%% file: 0_main.tex
\AtBeginDocument{%
  }
    
\documentclass[sigconf, 9pt]{acmart}
\usepackage{svg}
\usepackage{algorithmic}
\usepackage{graphicx}
\usepackage{textcomp}
\usepackage{xcolor}
\usepackage{hyperref}
\usepackage{soul}
\usepackage{physics,array}
\usepackage{multirow}
\usepackage{tikz}

\setcopyright{acmlicensed}
\copyrightyear{2018}
\acmYear{2018}
\acmDOI{XXXXXXX.XXXXXXX}

\renewcommand\footnotetextcopyrightpermission[1]{}

\acmBooktitle{Woodstock '18: ACM Symposium on Neural Gaze Detection,
 June 03--05, 2018, Woodstock, NY}
\acmISBN{978-1-4503-XXXX-X/18/06}

\begin{document}
\fancyhead{}

\title{Application of Quantum Tensor Networks for Protein Classification  }


\author{Debarshi Kundu}
\affiliation{%
  \institution{Pennsylvania State University}
  \city{State College}
  \state{PA}
  \country{USA}
  }
\email{dqk5620@psu.edu}

\author{Archisman Ghosh}

\orcid{0000-0002-0264-6687}
\affiliation{%
  \institution{Pennsylvania State University}
  \city{State College}
  \state{PA}
  \country{USA}
}
\email{apg6127@psu.edu}

\author{Srinivasan Ekambaram}

\orcid{}
\affiliation{%
  \institution{Pennsylvania State University}
  \city{State College}
  \state{PA}
  \country{USA}
}
\email{je5459@psu.edu}

\author{Jian Wang}

\orcid{}
\affiliation{%
  \institution{Pennsylvania State University}
  \city{State College}
  \state{PA}
  \country{USA}
}
\email{juw1179@psu.edu}

\author{Nikolay Dokholyan}

\orcid{}
\affiliation{%
  \institution{Pennsylvania State University}
  \city{State College}
  \state{PA}
  \country{USA}
}
\email{nxd338@psu.edu}

\author{Swaroop Ghosh}
\affiliation{%
  \institution{Pennsylvania State University}
  \city{State College}
  \state{PA}
  \country{USA}
}
\email{szg212@psu.edu}


\begin{abstract}
Computational methods in drug discovery significantly reduce both time and experimental costs. Nonetheless, certain computational tasks in drug discovery can be daunting with classical computing techniques which can be potentially overcome using quantum computing. A crucial task within this domain involves the functional classification of proteins. However, a challenge lies in adequately representing lengthy protein sequences given the limited number of qubits available in existing noisy quantum computers.
We show that protein sequences can be thought of as sentences in natural language processing and can be parsed using the existing Quantum Natural Language framework into parameterized quantum circuits of reasonable qubits, which can be trained to solve various protein-related machine-learning problems.
We classify proteins based on their sub-cellular locations—a pivotal task in bioinformatics that is key to understanding biological processes and disease mechanisms. Leveraging the quantum-enhanced processing capabilities, we demonstrate that Quantum Tensor Networks (QTN) can effectively handle the complexity and diversity of protein sequences.
We present a detailed methodology that adapts QTN architectures to the nuanced requirements of protein data, supported by comprehensive experimental results. We demonstrate two distinct QTNs, inspired by classical recurrent neural networks (RNN) and convolutional neural networks (CNN), to solve the binary classification task mentioned above. 
Our top-performing quantum model has achieved a 94\% accuracy rate, which is comparable to the performance of a classical model that uses the ESM2 protein language model embeddings. It's noteworthy that the ESM2 model is extremely large, containing 8 million parameters in its smallest configuration, whereas our best quantum model requires only around 800 parameters.
We demonstrate that these hybrid models exhibit promising performance, showcasing their potential to compete with classical models of similar complexity. 
\end{abstract}







\maketitle

\input{1_introduction}
\input{2_Model}

\input{3_Dataset}
\input{4_Methodology}
\input{5_Result}

\input{6_conclusion}

\begin{acks}
We extend our gratitude to Avimita Chatterjee (Pennsylvania State University) for her valuable insights and assistance in structuring the work. The work is supported in parts by the National Science Foundation (NSF) (CNS-1722557, CCF-1718474, OIA-2040667, DGE-1723687, and DGE-1821766).
\end{acks}

\bibliographystyle{unsrt}
\bibliography{refs}

\end{document}

%% file: 1_introduction.tex
\section{Introduction} \label{sec:introduction}
\textbf{Importance of Proteins in drug discovery: }
Proteins are essential, large biomolecules that serve numerous vital functions within the body.
Constructed from sequences of amino acids, proteins are the end product of long chains comprised of these units. The human body utilizes 20 distinct amino acids, the specific ordering of which crafts a protein's unique three-dimensional shape and determines its particular role. This order of amino acids is encoded by the DNA through sequences of genes, which specify the arrangement of three nucleotide building blocks to form each amino acid. Protein functions reflect their fundamental importance in biological processes. In the realm of drug discovery, the knowledge of protein structures and functions is crucial. For instance, understanding how a protein interacts with other molecules can guide the design of drugs that can modulate these interactions. A classic example is the development of inhibitors targeting HIV protease, a pivotal enzyme in the HIV life cycle. The structural and functional insights into the HIV protease have led to the creation of drugs that specifically inhibit this enzyme, significantly improving the treatment of HIV infection \cite{sudhakararao2019physiological}. This approach highlights the importance of protein sequence and structure knowledge in designing therapeutic agents that can effectively target specific molecular pathways \cite{anderson2004dietary}.

\textbf{Machine learning for protein engineering:}
The incorporation of computational methods, notably in protein design, has significantly shortened experimental timelines in drug discovery. The advent of machine learning (ML) has particularly transformed protein sequence analysis, as evidenced by innovations like ESM-2\cite{lin2023evolutionary} and AlphaFold\cite{bryant2022improved}. ESM-2 has improved our predictive capabilities for protein functions from sequences, while AlphaFold has revolutionized protein structure prediction, reaching near-experimental accuracy. AlphaFold's success in the CASP competitions highlights this progress. These advancements underscore the efficiency of ML in enhancing our comprehension of protein structures and functions, leveraging large datasets to reveal previously unknown predictive relationships. Despite these achievements, the complexity of some computational biology problems exceeds the capabilities of classical computing, suggesting a pivotal role for quantum machine learning in navigating these challenges more effectively.

\begin{figure}
    \centering
    \includegraphics[width=0.7\linewidth]{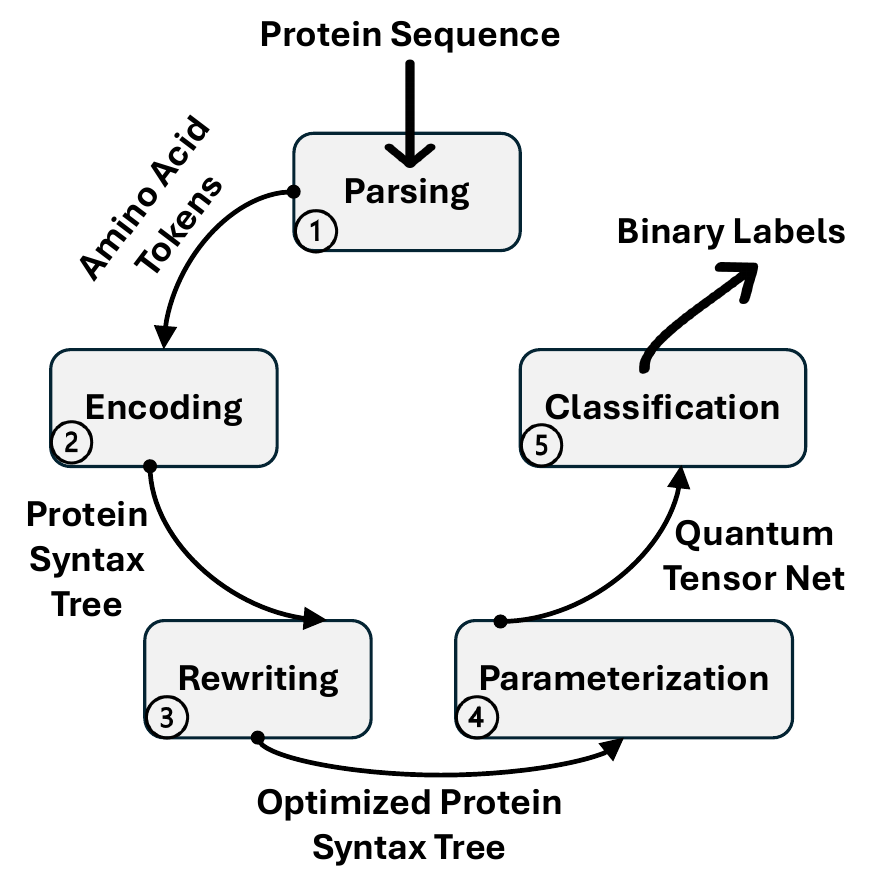}
    \caption{\textbf{A diagram describing the flow of binary classification of the protein sequence.
     } }
    \label{fig:flow}
    \vspace{-10pt}
\end{figure}

\textbf{Quantum Natural Language Processing (QNLP):} The application of quantum computing to Natural Language Processing (NLP) involves representing word embeddings as quantum states in quantum Hilbert space to obtain a polynomial time speedup in classification-based problems. The word embeddings in QNLP are prepared using linear maps to the tensor product of the word vectors, to find the meaning of a sentence \cite{coecke2010mathematical}. PQCs are the primary trainable component in the QNLP circuit comprising entangling operations between qubits and parameterized single-qubit rotations. The entanglement operations are mostly multi-qubit operations between all qubits to generate correlated states and the parameterized rotations are used to search the solution space. The QNLP circuit is a Quantum Tensor Network that represents sentences as high-dimensional tensors and parses them into a network of more manageable, lower-dimension tensors that aid classification.

\textbf{Protein Sequence as a sentence:} 
Viewing a protein sequence as a sentence, with each amino acid acting as a word, presents a powerful analogy for deciphering the structure and function of proteins \cite{ofer2021language}. This perspective highlights the complexity and specificity inherent in protein sequences, likening a protein to a carefully composed sentence where the amino acids are arranged in a precise sequence. Each "word" (amino acid) adds its unique characteristics to the "sentence" (protein sequence), influencing the protein's folding, structure, and function within biological systems. For example, the positioning of amino acids such as lysine, arginine, and glutamate within a protein can be seen as words constructing a sentence, where the exact sequence and arrangement are critical for the protein's capability to fulfill its designated roles. Similarly, altering a word in a sentence can change its entire meaning, just as modifying an amino acid in a protein sequence can drastically affect the protein’s functionality. This analogy emphasizes the critical nature of sequence fidelity in proteins, underscoring the finely tuned equilibrium of biological systems where each "word" plays a pivotal role in the "narrative" of life \cite{yandell2002genomics}.

\textbf{Motivation:}
Integrating Quantum Natural Language Processing (QNLP) into protein sequence analysis offers the potential to revolutionize drug discovery and our understanding of biological processes. This approach leverages quantum computing's capacity for high-dimensional space management, speed, and semantic analysis, enabling more accurate predictions of protein functions, structures, and interactions. By viewing amino acid sequences as sentences, Quantum NLP can provide deeper insights into the protein "language," surpassing current bioinformatic tools in sequence alignment, functional motif identification, and protein function annotation\cite{khatami2023gate}, \cite{hollenberg2000fast}. This integration could significantly advance drug discovery by exploiting quantum algorithms for enhanced efficiency and speed. However, there is a need to represent the long protein sequences in the quantum circuits with reasonably small qubits and reasonably deep quantum circuits in the NISQ-era quantum computers. There is a need to extract the signal from these sequences to solve important challenges in drug discovery. 

\textbf{Contributions:}
Building on the foundational work by \cite{harvey2023sequence}, our study marks the proof of concept of classifying long protein sequences leveraging Quantum Tensor Networks (QTNs). This advancement is pivotal, as it not only extends the applicability of QTNs beyond their traditional domains but also introduces a novel methodology for handling the complexities inherent in protein sequence data. The utilization of quantum computing in this context is not merely for its computational prowess but also for its ability to capture the intricate patterns and relationships within biological sequences, which are often beyond the reach of classical computational techniques.


\begin{figure}
    \centering
    \includegraphics[width=0.7\linewidth]{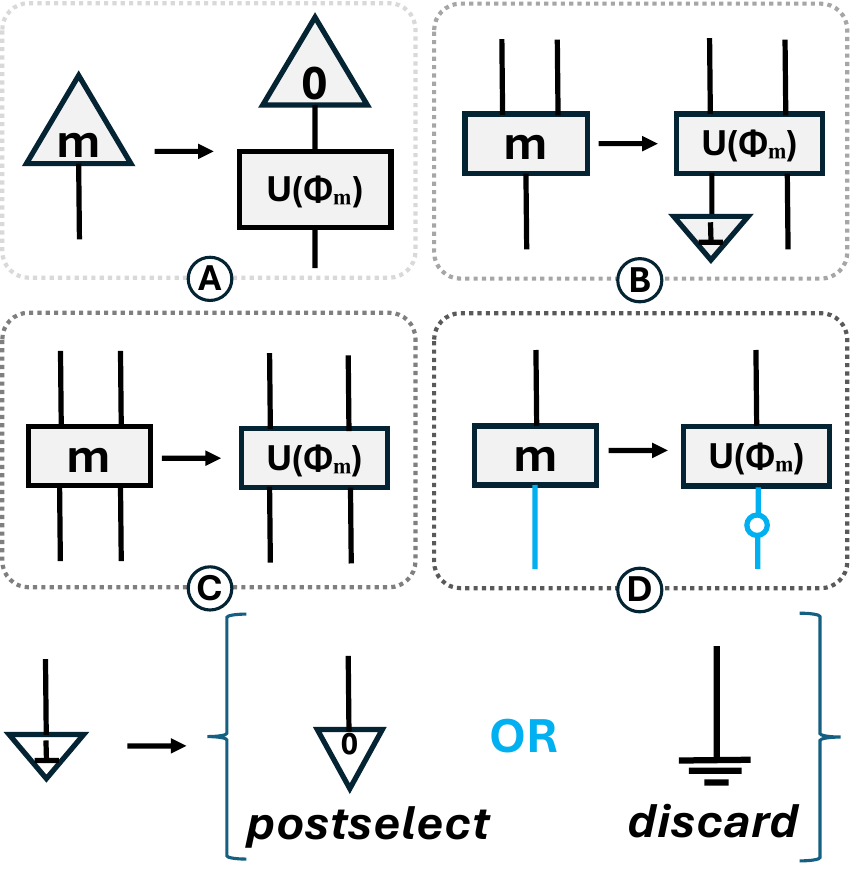}
    \vspace{-10pt}
    \caption{Assignment of parameterized quantum circuits
$U(\phi_i)$ to boxes labeled $i$: This is an example of a function
definition, where the $words$ are mapped to qubits (bits in case
of blue wires) via the unitary matrices $U(\phi)$. The $\perp$ is either represented as an all-zeroes state, postselect, or discard.}
    \label{fig:schematic}
    \vspace{-15pt}
\end{figure}

In a nutshell, our contributions to this paper are as follows:
(a) we have successfully demonstrated, for the first time, the potential of QTNs\cite{harvey2023sequence} in the classification of long protein sequences, 
(b) we showed that QTNs, inspired by convolutional and recurrent neural networks, are capable of learning representations of proteins using a relatively small qubit circuit and (c)
our findings underscore a significant advancement over classical models, showcasing the inherent advantages of quantum computing in processing and classifying biological data. 
This comparison not only validates the effectiveness of our approach but also sets the stage for future explorations into quantum bioinformatics.



%% file: 2_Model.tex
\section{Proposed Models} \label{sec:background}
Here we first explain the entire process of generating a parametrized quantum circuit from a protein sequence and training the circuits for the binary classification task (Fig. \ref{fig:flow}). Then we explain two important steps to develop the quantum models namely, (a) building compositional schemes to convert protein sequences into networks which are represented using wires and boxes and (b) defining semantic functor to map these networks into quantum circuits. 
\\
\textbf{Protein to quantum model pipeline:}
The process begins with a protein sequence input that undergoes parsing using a state-of-the-art neural-trained parser, resulting in a protein syntax tree. This tree is encoded into a string diagram, abstractly representing the relationships between elements in the sequence. These string diagrams are based on category theory \cite{pregrp} and can be simplified by rewriting rules to reduce redundancy and adapt the computation for quantum processors. After rewriting, the diagrams are parameterized and converted into a quantum circuit using specific parameterization schemes and ansätze choices. The quantum circuit is then classified using a QTN, resulting in binary labels, and is ready for training. This entire model is structured based on the grammatical construction of the input sequence and is optimized for implementation on quantum processing units.


\begin{figure}
    \centering
    \includegraphics[width=0.7\linewidth]{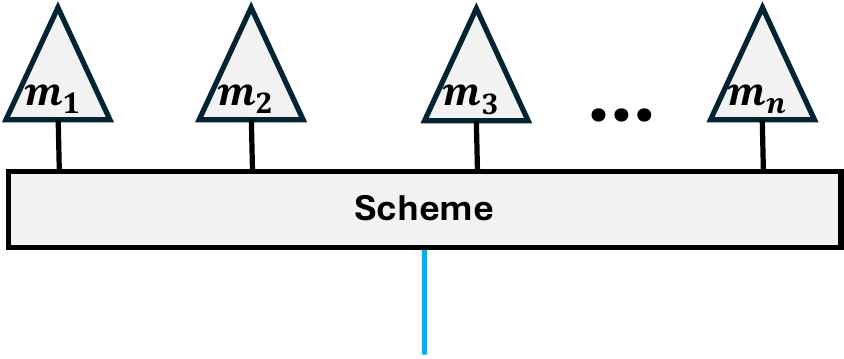}
    \vspace{-10pt}
    \caption{A sentence is broken into $words$ $m_1...m_n$, converted to corresponding unitaries based on the Functor rules in Fig. \ref{fig:schematic} and finally run through the Scheme ($QTN$). }
    \label{fig:scheme}
    \vspace{-10pt}
\end{figure}

\textbf{Compositional Schemes:}
A compositional scheme is initially defined for a given sequence, employing the graphical language of process theories. The processes within these schemes are represented by boxes, which have input and output wires. These wires carry types - either the 'internal' type $\tau$ or the 'sentence' type $\sigma$. The composition of these boxes, following type constraints, allows for the generation of process diagrams that represent the scheme for sequence analysis. Given a vocabulary $\mathcal{V} = \{m_i\}_{i}$ comprising a finite set of words (or tokens), we consider compositional schemes for sequences $\mathcal{S}$ of finite length over this vocabulary. The schemes are then semantically mapped onto QTN models (Fig.  \ref{fig:schematic}).

\textbf{Semantic Functor $\mathcal{F}$:} It is a structure-preserving map that assigns Hilbert space semantics to the compositional schemes, thereby designing parameterized quantum circuits (PQCs) for various components of the scheme. For handling non-deterministic outcomes, two strategies are defined: \texttt{postselect} and \texttt{discard}. \texttt{Postselect} involves conditioning on a particular measurement outcome, typically the all-zeros state, while \texttt{discard} involves ignoring specific dimensions of the quantum state like a partial trace.

\textbf{Parameterized Quantum Circuits (PQCs) for QTNs:}
The PQCs designed by $\mathcal{F}$ involves \textbf{Word-State Preparation} (Fig. \ref{fig:schematic}A):
These boxes prepare a word-state of type $\tau^{\otimes 0} \rightarrow \tau^{\otimes 1}$, which is associated with a parameterized quantum state prepared by applying the circuit $U(\phi_m)$ to the fixed input state $|0\rangle^{\otimes q}$. Each word corresponds to a unique set of parameters $\phi_m$,
\textbf{Filter Application} (Fig. \ref{fig:schematic}B): The filter boxes, with type $\tau^{\otimes 2} \rightarrow \tau^{\otimes 2}$, are associated with a $U(\phi_m)$ operating on $2^q$ input and $2^q$ output qubits, simulating the filtering process within the protein sequence,
\textbf{Merge Operation }(Fig. \ref{fig:schematic}C): The m-box, typed $\tau^{\otimes 2} \rightarrow \tau^{\otimes 1}$, is mapped to $U(\phi_m)$, with $2^q$ input qubits and $q$ output qubits. The reduction in qubits is achieved by either discarding or postselecting the redundant qubits via the $\perp$-effect, and 
\textbf{Classification }(Fig. \ref{fig:schematic}D): The classifier boxes, typed $\tau^{\otimes 1} \rightarrow \sigma^{\otimes 1}$, is associated with a unitary $U(\phi_m)$, which processes a $q$-qubit state input and outputs a $q'$-qubit state. This state is subsequently measured in the Z basis, yielding a vector in $[0,1]^{2^{q'}}$, representing the classification outcome based on the Born rule probabilities. These strategies influence the tensor network topology, allowing for efficient tensor contraction which encapsulates the sequence processing task within a quantum framework.

\begin{figure}
    \centering
    \includegraphics[width=0.9\linewidth]{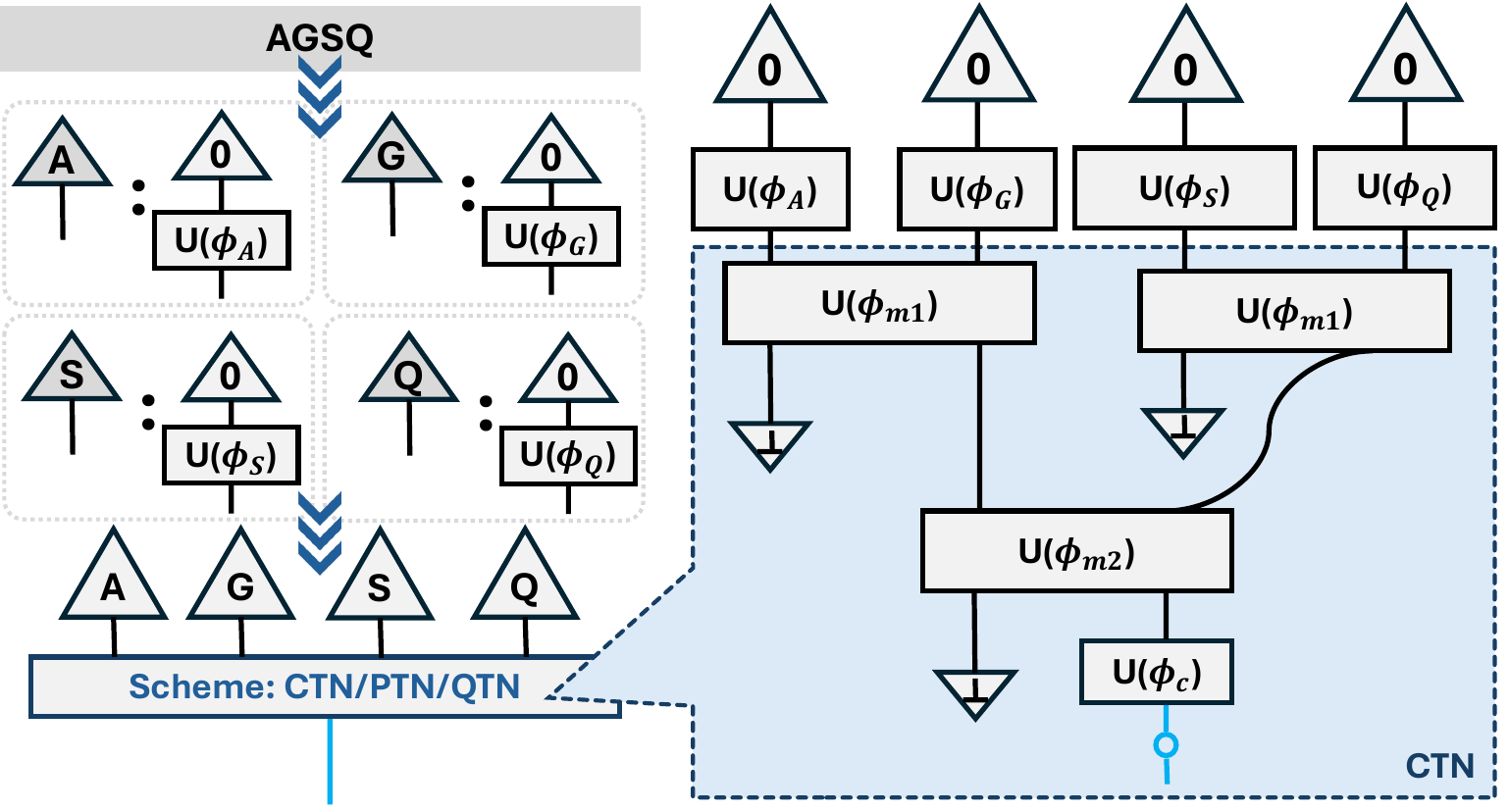}
    \vspace{-5pt}
    \caption{\textbf{An example to parse an demonstrative protein sequence $AGSQ$ into a protein syntax tree based on CTN.} }
    \label{fig:example}
    \vspace{-5pt}
\end{figure}

\begin{figure}
    \centering
    \includegraphics[width=0.7\linewidth]{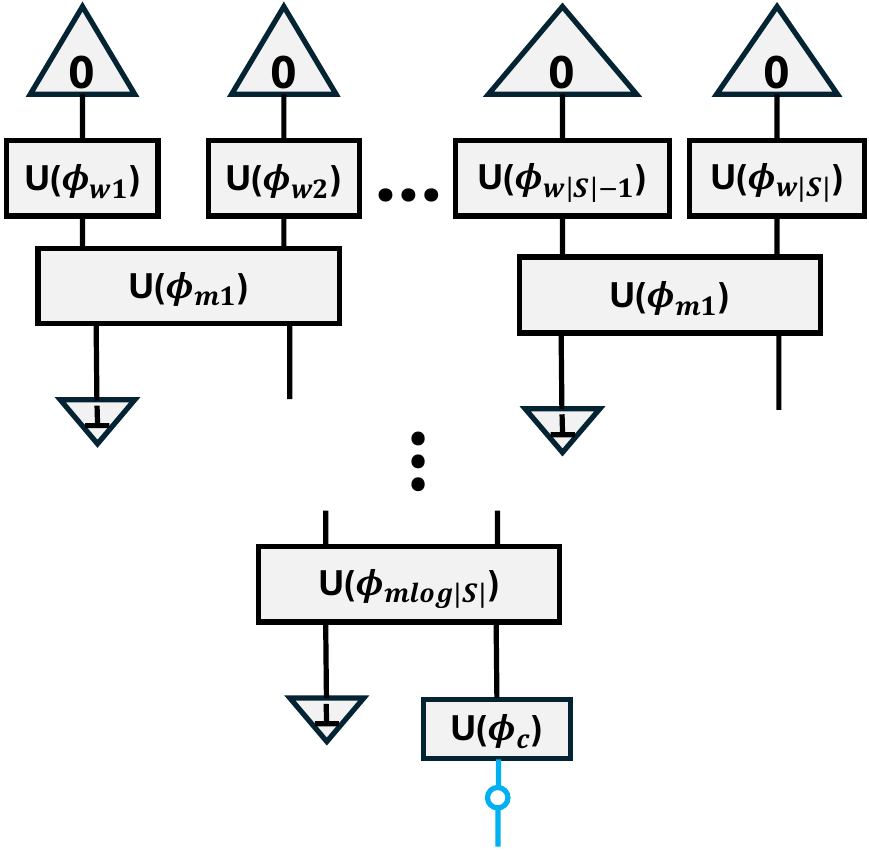}
    \vspace{-8pt}
    \caption{\textbf{Convolutional tensor network (CTN)} }
    \label{fig:CTN}
    \vspace{-10pt}
\end{figure}

\textbf{Example:} A protein sequence is represented as a sentence and broken down into independent $words$ and then run through the QTN (Fig. \ref{fig:scheme}). We dive deep into the idea by considering a protein sequence of length four ($AGSQ$) in Fig. \ref{fig:example} and define the $amino \space acids$ as different $words$ that are then converted to the respective unitaries (parameterized input states) based on the rules of the Functor mapping (Fig. \ref{fig:schematic}). These mapped states finally culminate into the Quantum Convolution Tensor Net (Fig. \ref{fig:CTN}).

\subsection{Compositional scheme: Recurrent Neural Net inspired}
\begin{figure}
    \centering
    \includegraphics[width=0.6\linewidth]{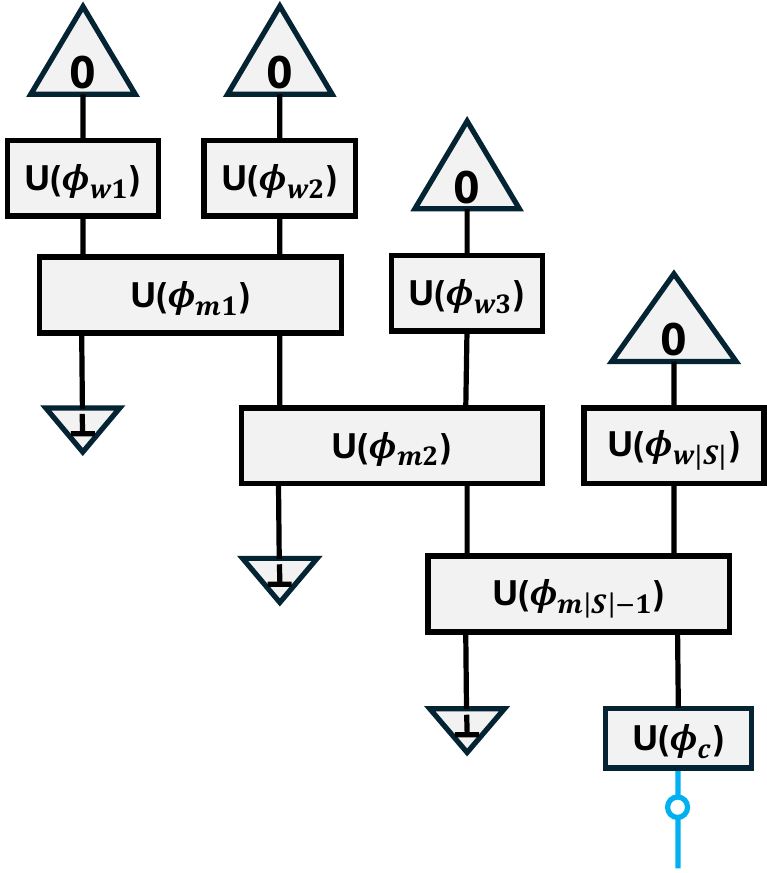}
    \vspace{-5pt}
    \caption{\textbf{Path tensor network (PTN)} }
    \label{fig:PTN}
    \vspace{-10pt}
\end{figure}

We examine QTNs with a model that adheres to the sequential flow akin to the natural progression of words in a text. This model, derived by implementing the semantic functor $\mathcal{F}$ outlined in Fig. \ref{fig:schematic}, results in what we refer to as the path tensor network model, symbolized as $\mathcal{F}(\phi)_{\text{[path]}} = \text{PTN}$, and illustrated in Fig. \ref{fig:PTN}. This approach sequentially aligns with the reading order, mapping out a straightforward path through the sequence.

\subsection{Convolutional Tensor Networks}
\textbf{Hierarchical and Uniform PTN Models:}
Delving deeper, we introduce a nuanced layer to the model by attributing a hierarchical structure to the parameter sets $\{\phi_{mi}\}$, contingent upon their sequential position $i$, within the range of $\{1,2,\ldots,|S|-1\}$. This adjustment births the hierarchical PTN (hPTN) models. Proceeding further, we harmonize the parameters across all merging circuits ($m$-circuits) to a singular set, $\phi_m = \phi_{mi}$ for all $i$, thereby engendering a recurrent structure, which we term the uniform PTN (uPTN). This uPTN model distills down to a basic form of a recurrent quantum model, or equivalently, a matrix product state (MPS) model, in which all dimensions except for the last are either disregarded or selected based on outcomes, with the remaining dimension capturing the overall semantic value of the sentence. This initial compositional approach intentionally bypasses considerations of syntax and distant correlations, positioning it as an elementary framework for juxtaposition with models that incorporate syntactic awareness. While simplistic, this model is built upon a principle of local compositional application, serving as a critical benchmark.

Our approach uses a refined compositional scheme, similar to convolutional neural networks, which serves as an advancement over the basic tree structure. This scheme is ingeniously crafted by integrating additional layers of filtering boxes (f-boxes) into the tree architecture. These f-boxes are strategically placed to operate prior to the merging boxes (m-boxes) along adjacent wires that do not converge into the same m-box. This setup results in a convolutional tensor network (CTN), distinguished by its ability to filter out superfluous entanglements at each layer through the f-circuits, followed by a consolidation of qubit wires by the m-circuits. This process effectively distills the sequence, preserving only the essential information pertinent to the designated task.
\textbf{Hierarchical and Uniform Variants:}
The CTN model evolves into hierarchical (hCTN) and uniform (uCTN) variants based on the distribution and uniformity of the parameter sets across the layers. The hierarchical model shares parameter sets within the same layer, whereas the uniform model extends this sharing across the entire model. 

\subsection{Classical Model}
\begin{figure}[t]
    \centering
    \includegraphics[width=0.8\linewidth]{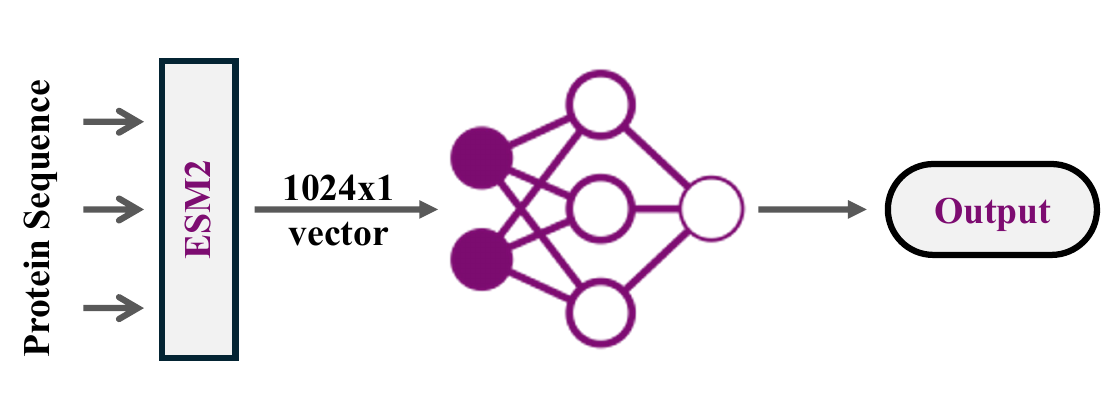}
    \vspace{-10pt}
    \caption{Protein sequence is fed into the ESM2 embedding that passes through a fully connected classical layer to classify the input sequence.}
    \label{fig:classical_model}
    \vspace{-10pt}
\end{figure}

To establish a baseline for rigorously assessing the capabilities of our quantum models, we developed a classical model architecture (Fig. \ref{fig:classical_model}) that exploits deep learning techniques, specifically tailored for processing and classifying protein sequences. Central to our classical model is the integration of embeddings derived from the ESM2 \cite{lin2023evolutionary} pretrained model. Esteemed as a cutting-edge development in machine learning for bioinformatics, ESM2 is intricately designed to distill meaningful features from protein sequences, thereby representing a substantial leap forward in our ability to capture the complex patterns and functional attributes inherent in proteins. This model's ability to learn from an extensive compendium of protein sequences endows it with the capacity to abstract a profound representation of amino acid interrelations, structural motifs, and other critical biochemical properties.

Within our classical framework, each protein sequence undergoes initial processing by the ESM2 model, yielding a fixed-size (1024) embedding vector. This vector serves as a condensed representation of the sequence's biological and contextual nuances, primed for subsequent analysis via neural network techniques. These embeddings are then channeled into a fully connected network, comprising three hidden layers of size ( 512, 256, 128) of interconnected neurons. 
The training regimen for this model utilizes the same labeled dataset of protein sequences mentioned earlier, employing binary cross-entropy as the loss function to fine-tune the network's weights. Adam, renowned for its optimization efficacy \cite{kingma2014adam}, was the algorithm of choice for this process, facilitating efficient and effective model training.
By juxtaposing this classical model's performance against that of our quantum approaches, we endeavor to elucidate the quantum computing paradigm's potential benefits and efficacy in tackling the challenges associated with protein sequence classification.

%% file: 4_Methodology.tex
\section{Methodology}
\subsection{Dataset}
\textbf{Compilation:}
The dataset has been compiled using protein sequences obtained from UniProt \cite{uniprot2023uniprot}. The dataset of human protein sequences has been cleaned and preprocessed to have 80 to 200 amino acids in each sequence. The protein sequences have a categorization that has been based on subcellular localization into two groups—proteins in the cytosol or cytoplasm and those associated with the cell membrane. \textbf{Structure:} 
The dataset is made of 1136 protein sequences after preprocessing which is divided into training, validation, and testing data subsets having 980, 123, and 123 protein sequences respectively. The size of the dataset has been scaled down to cater to the long runtime of quantum simulations while ensuring model validation and a thorough evaluation of its predictive performance. \textbf{Significance:} The dataset comprises protein sequences that can be classified based on their location in the cell (cytoplasm or cell membrane), therefore facilitating a deeper understanding of the functional implications of proteins based on their cellular locales. On a broader scale, it contributes to the study of complex cellular functions, and biological processes.
\\

\subsection{Implementation Details}
The \texttt{lambeq} \cite{kartsaklis2021lambeq} library, a forefront tool in quantum natural language processing (QNLP), introduces a sophisticated method for translating diagrammatic representations of linguistic structures into quantum circuits, enabling the exploration of various quantum ansatzes for processing and analysis. The ansatzes explored include the \texttt{IQPAnsatz}, \texttt{Sim14Ansatz}, \texttt{Sim15Ansatz}, and \texttt{MPSAnsatz}, each offering unique advantages for different types of quantum computations~\cite{sim2019expressibility} as explained next.
\textbf{IQPAnsatz (Instantaneous Quantum Polynomial-time Ansatz):}
This ansatz constructs circuits that are believed to implement computations not efficiently simulatable by classical computers, focusing on problems that can be encoded in a certain polynomial structure. It is particularly suited for tasks where quantum advantage is explored.

The mapping from diagrammatic representations to quantum circuits within \texttt{lambeq} is facilitated by parsers like \texttt{spiders\_reader}, \texttt{cups\_reader}, and \texttt{stairs\_reader} \cite{kartsaklis2021lambeq}. \texttt{spiders\_reader} translates complex syntactic interactions into quantum circuits using graphical elements known as spiders, facilitating the representation of non-linear word relationships. \texttt{cups\_reader} captures entanglement between elements in a sentence through cups in diagrammatic notation, effectively modeling pairwise dependencies. \texttt{stairs\_reader} leverages a staircase pattern to represent the sequential flow and dependencies of elements within a sentence, ideal for capturing long-range contextual information crucial in understanding protein sequences.

The transition from these diagrammatic representations to quantum circuits involves encoding linguistic or biological data as initial quantum states, followed by the application of quantum gates as dictated by the chosen ansatz. This process effectively translates the structure and semantics of the input data into a form amenable to quantum computation, enabling the exploration of quantum mechanical advantages in processing complex sequences. Simulation and training of these models are facilitated through the \texttt{tensornetwork} library and JAX, respectively, with the latter enabling Just-In-Time compilation for efficient processing \cite{roberts2019tensornetwork} \cite{jax2018github}. Among various quantum ansatz (\texttt{IQPAnsatz}, \texttt{Sim14Ansatz}, \texttt{Sim15Ansatz} \cite{sim2019expressibility}, \texttt{MPSAnsatz}), the expressive \texttt{ansatz 14} was selected for its notable test performance \cite{sim2019expressibility}. Optimization is achieved using AdamW, with the aim of minimizing binary cross-entropy loss for accurate label prediction~\cite{AdamW}. Considering the prospect of quantum computer training, we suggest the parameter-shift rule for gradient estimation or the use of SPSA for its practicality in near-term quantum computing environments~\cite{ParameterShift2019, SPSA119632, PerformanceComparisonOptimisers2023}.

The model selection process uses k-fold validation and early stopping, focusing on hyperparameters like embedding qubit count, ansatz depth (q, D), and learning rate. In k-fold validation, the data is split into k parts, training on k-1 and validating on the remaining part, iteratively. This ensures a comprehensive evaluation. The model's generalization is finally tested on an unseen dataset, assessing prediction accuracy.
We ensure a consistent comparison framework by reporting test accuracies at the peak of validation accuracy for specified hyperparameters and learning rate settings, under a fixed seed for reproducibility.

%% file: 5_Result.tex
\section{Results} \label{sec:evaluation}

Our model follows an encoder architecture, which essentially means they accept sentences as inputs and generate corresponding outputs, thereby serving the role of classifiers. These models are predominantly aimed at binary classification tasks, involving the measurement of a singular qubit ($q' = 1$) from the output quantum state produced by the $U_{c}$ circuit. This process determines probabilities for two possible outcomes, $p_0$ and $p_1$ by measuring the average state of the qubits, with each outcome directly mapping to a class label. The tree-like design of our introduced model species promotes not only efficient computation but also a natural resistance to the occurrence of barren plateaus during the training process, a notable hurdle in the optimization of quantum models\cite{noBPs,Enrique2023, zhao2021analyzing}.

\begin{table}[]
\caption{Test accuracy for evaluated models under different measurement procedures (\texttt{discard}) and (\texttt{postselect})}
\begin{tabular}{c|cccc}
           & \textbf{uPTN} & \textbf{hPTN} & \textbf{uCTN} & \textbf{hCTN} \\ \hline
\textbf{discard}    & 0.73 & 0.94 & 0.57    & 0.71    \\ \hline
\textbf{postselect} & 0.72 & 0.83 & 0.60 & 0.67 \\ \hline
\end{tabular}
\label{tab:accuracy_table}
\vspace{-10pt}
\end{table}

\begin{table}[]
\caption{F1-score for evaluated models under different measurement procedures (\texttt{discard}) and (\texttt{postselect})}
\begin{tabular}{c|cccc}
           & \textbf{uPTN} & \textbf{hPTN} & \textbf{uCTN} & \textbf{hCTN} \\ \hline
\textbf{discard}    & 0.86 & 0.97 & 0.71    & 0.82    \\ \hline
\textbf{postselect} & 0.83 & 0.84 & 0.74 & 0.80 \\ \hline
\end{tabular}
\label{tab:f1_table}
\vspace{-10pt}
\end{table}

\begin{figure*}
    \centering
    \includegraphics[width=0.8\linewidth]{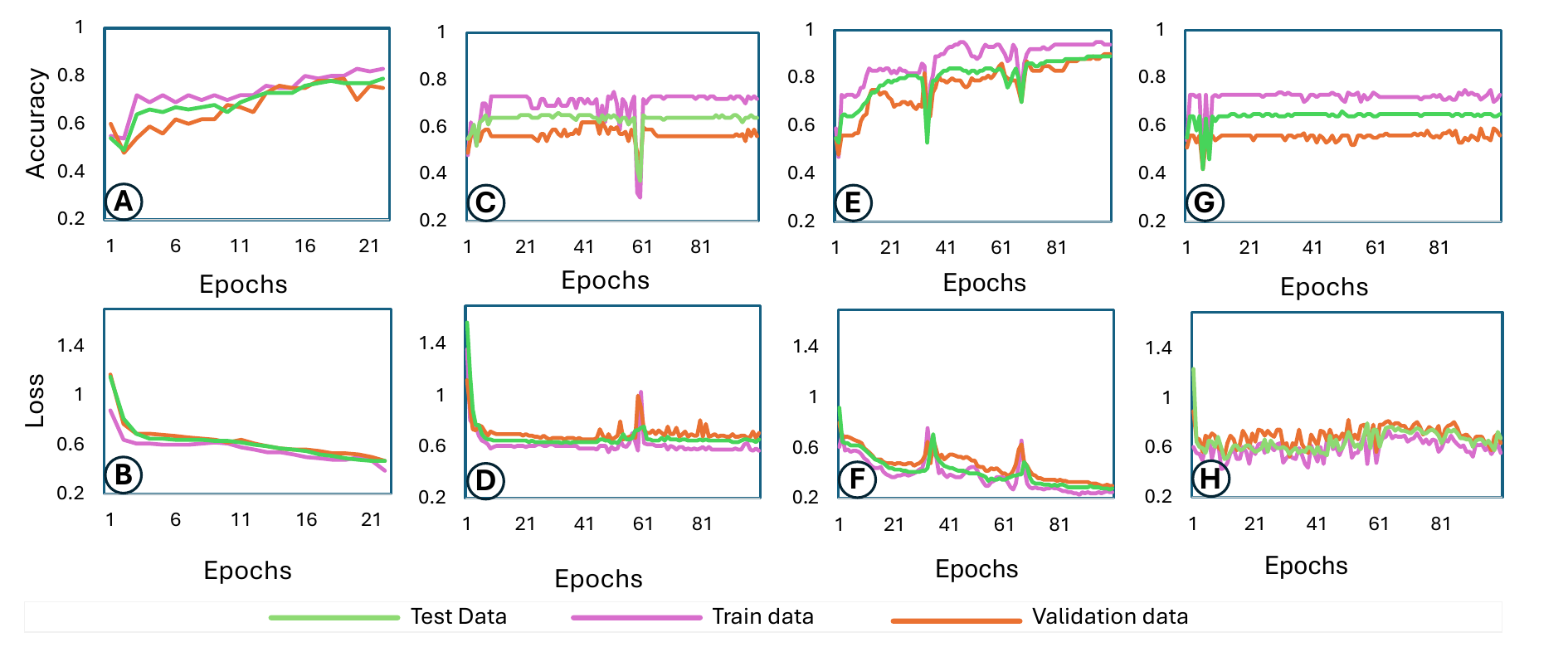}
    \vspace{-5pt}
    \caption{\textbf{Accuracy and Loss plots for training, test, and validation data evaluated for PTN models under \texttt{postselect} and \texttt{discard} measurements. Plots A-D represent \texttt{postselect} measurement for hierarchical (A, B) and uniform (C, D) PTNs. Plots E-H represent \texttt{discard} measurement for hierarchical (E, F) and uniform (G, H) PTNs.} }
    \label{fig:loss_accuracy}
    \vspace{-10pt}
\end{figure*}
We have explored different Quantum Tensor Networks, viz., the Path Tensor Network (PTN) and the Convolutional Tensor Network (CTN)—with hierarchical and uniform parameter-sharing strategies. We analyze the performance of four models: the uniform path tensor network (uPTN), the hierarchical path tensor network (hPTN), the uniform convolutional tensor network (uCTN), and the hierarchical convolutional tensor network (hCTN).

Our results, summarized in two tables (Table \ref{tab:accuracy_table} and \ref{tab:f1_table}), provide the test accuracy and F1-scores for each model, with and without post-selection. The hPTN model outperforms the others in both metrics significantly, indicating its superior capability in capturing the necessary features for classification. The uPTN follows, showing decent performance, but not matching the hierarchical counterpart. The uCTN and hCTN models exhibit lower performance, with uCTN scoring the lowest on both accuracy and F1-score.

\textbf{Model Performance Analysis:}
The PTN-based models have a sequential architecture that aligns well with the natural sequence of amino acids in protein sequences allowing better classification accuracy. In \textbf{hPTN} models (Fig. \ref{fig:loss_accuracy}), the sequential information of protein sequences along with the varying dependencies are captured very effectively due to the presence of unique parameter sets in the model. \textbf{uPTN} models, based on recurrent quantum nets, perform consistently well but are limited in flexibility by a single shared parameter set across all PQCs. Although \textbf{CTNs} have enhanced tree-like architecture, they are not very adept at capturing the long-range correlations effectively. The slight improvement of the performance of \textbf{hCTN} models over \textbf{uCTN} models suggests that having a hierarchical unique parameter set helps in classification more than having a single shared parameter set.
In the comparative analysis of quantum models for protein sequence classification, the ESM2-based classical model (Fig. \ref{fig:classical_model}), with its high accuracy of 0.98, sets a significant benchmark for performance. Notably, the hierarchical Path Tensor Network (hPTN) quantum model closely rivals this benchmark with an impressive accuracy of 0.94. This near-parity highlights the substantial potential of quantum models to reach and possibly exceed the performance standards set by advanced classical models in complex biological computations.

\textbf{Model runtime analysis:} 
CTN models, characterized by their substantial parameter count, require significantly longer training durations, approximately 8 hours per epoch. Conversely, PTN models exhibit a markedly swifter training pace, completing the entire simulation in about 1 hour. Intriguingly, CTN models achieve convergence in fewer epochs (around 4 to 5) compared to PTN which took much higher number of epochs as shown in Fig. \ref{fig:loss_accuracy}. Despite the classical model being trained within an hour, it's important to acknowledge that the foundational pre-trained protein language model, ESM2, underwent a training process powered by extensive computational resources for nearly a week.

\textbf{Limitations:} The QTNs were evaluated under idealized conditions devoid of quantum noise, likely leading to an optimistic representation of their capabilities. Such an omission of quantum noise considerations prevents a fully equitable comparison to the classical ESM2 model and overlooks the challenges posed by the hardware noise in the devices of the current noisy quantum computers. Future research should prioritize addressing this gap by incorporating noise models in the simulation backends. \\

%% file: 6_conclusion.tex
\vspace{-22pt}
\section{Conclusion} 
\label{sec:conclusion}


We proposed and evaluated four flavors of Quantum Tensor Nets (QTNs) namely, hierarchical Path Tensor Network (hPTN), uniform Path Tensor Network (uPTN), hierarchical Convolutional Tensor Network (hCTN), and uniform Convolutional Tensor Network (uCTN) to classify protein sequences based on their cellular locales (cytosol/cytoplasm or cell membrane). 
The hPTN model demonstrated superior performance in classifying protein sequences compared to its uniform counterpart and the convolutional tensor network variants. The uniform models, particularly the uCTN, require further investigation and possible architectural or parameter adjustments to improve their performance. 
